# Signatures of Helical Edge Transport in Millimetre-Scale Thin Films of $Na_3Bi$

Chang Liu[1,2,3], Dimitrie Culcer[3,4], Mark T. Edmonds[1,2,3*], Michael S. Fuhrer[1,2,3*]

[1]School of Physics and Astronomy, Monash University, Victoria 3800 Australia

[2]Monash Centre for Atomically Thin Materials, Monash University, Victoria 3800 Australia

[3]ARC Centre of Excellence in Future Low-Energy Electronics Technologies, Monash University, Victoria 3800 Australia

[4]School of Physics, University of New South Wales, Sydney, New South Wales 2052 Australia

\* mark.edmonds@monash.edu and michael.fuhrer@monash.edu

**Abstract**

A two-dimensional topological insulator (2DTI) has an insulating bulk and helical spin-polarised edge modes robust to backscattering by non-magnetic disorder. While ballistic transport has been demonstrated in 2DTIs over short distances, larger samples show significant backscattering and a nearly temperature-independent resistance whose origin is unclear. 2DTI edges have shown a spin polarisation, however the degree of helicity is difficult to quantify from spin measurements. Here, we study 2DTI few-layer $Na_3Bi$ on insulating $Al_2O_3$. A non-local conductance measurement geometry enables sensitive detection of the edge conductance in the topological regime, with an edge mean free path ~100 nm. Magnetic field suppresses spin-flip scattering in the helical edges, resulting in a giant negative magnetoresistance (GNMR), up to 80% at 0.9 T. Comparison to theory indicates >98% of scattering is helical spin scattering significantly exceeding the maximum (67%) expected for a non-helical metal. GNMR, coupled with non-local measurements demonstrating edge conduction, thus provides an unambiguous experimental signature of helical edges that we expect to be generically useful in understanding 2DTIs.

When time-reversal symmetry is preserved, the edges of a 2DTI are expected to be perfect ballistic conductors at zero temperature [1]. Indeed, 2DTIs were first identified experimentally through ballistic edge-state transport [2,3] over few-micron distances. However, in larger samples a finite, nearly temperature-independent resistance is observed [1,4-13]. The prevailing theoretical picture is that this edge resistance results from spin-flip scattering via exchange coupling to local moments, either impurity species [1] or metallic "puddles" of conduction electrons with odd electron number [8], however no experiments have probed this magnetic scattering directly.

Recent efforts have focused on identifying larger bandgap 2DTIs with the hope of demonstrating topological electronic devices at room temperature. WTe$_2$ [12], bismuthene [14] and Na$_3$Bi [15] are thought to be large-bandgap 2DTIs, edge-state transport has been shown up to 100 K in WTe$_2$ [11,13]. However, HgTe quantum wells have shown the persistence of edge state conduction into the conventional insulator regime, calling into question whether edge conduction alone is decisive evidence of topological transport. As the study of 2DTIs expands to new materials and new regimes, an unambiguous signature of helical edge state transport is necessary.

Here we demonstrate such a signature, using a combination of non-local transport measurements to demonstrate edge conduction, and magnetoresistance to confirm unambiguously helical transport. We study ultra-thin Na$_3$Bi, especially promising for topological devices due to its very large topological bandgap (>300 meV) and electric-field induced topological phase transition [15]. Unlike monolayer WTe$_2$ and bismuthene, Na$_3$Bi is expected to be a 2DTI over at least 1-3 layer thicknesses. Na$_3$Bi can be grown on a variety of substrates [16,17] at relatively low substrate temperature (≤ 330 °C). These qualities are promising for large area epitaxial ultra-thin 2DTIs, robust to temperature and layer-number fluctuations [15].

We grow large area epitaxial ultra-thin (~2nm) film Na$_3$Bi on an atomically flat insulator (α-Al$_2$O$_3$[0001]). We use local and non-local [4,10] conductivity measurements to probe the bulk and edge conductivity as a function of doping and magnetic field. As-grown bulk-conducting films possess carrier mobility as high as 34,000 cm$^2$/Vs and electron doping as low as $2.8 \times 10^{10}$ cm$^{-2}$, and show a carrier density-dependent transition between weak localization (WL) near the band edge to weak anti-localization (WAL) at high doping, characteristic of a gapped Dirac system. When the Fermi level is tuned into the bulk gap the non-local resistance becomes comparable to the local resistance, indicating edge-dominated transport even in our millimetre-sized devices. An exceptional giant negative magnetoresistance (GNMR), up to 80% at 0.9 T, develops as a direct consequence of the spin-flip nature of scattering in a topological edge mode. Comparison with theory indicates that >98% of backscattering is spin-flip scattering, as expected for the helical edges of a 2DTI, and inconsistent with a non-helical metal.

The Na$_3$Bi films were grown via molecular beam epitaxy (MBE) in a Createc Low-Temperature-MBE-STM system on atomically flat α-Al$_2$O$_3$[0001] substrates (Shinkosha Japan) that were prepatterned with Ti/Au electrodes (5/50 nm) in either a van der Pauw geometry or in a Hall bar geometry by growing through a stencil mask affixed on the substrate. The growth was achieved by co-depositing Bi (99.999%) and Na (99.95%) in an over Na flux with a Na:Bi flux ratio above 10:1. Deposition rates were calibrated using a quartz crystal microbalance. A two-step growth method was employed with the first 1 nm grown at 200 ℃ and the remaining 1 nm grown at final temperature around 270 ℃. After growth, annealing at 300-330 ℃ for 10 min in a Na flux was carried out to improve film quality. The as-grown film was then transferred in UHV to the low-temperature STM with 1 T perpendicular magnetic field and in-situ electrical contacts. The tetrafluorotetracyanoquinodimethane (F4-TCNQ) deposition was carried out in the MBE chamber at room temperature using an effusion cell and the rate calibrated by quartz crystal microbalance to *p*-type dope thin-film Na$_3$Bi without degrading the electronic properties [16,18]. The thickness of a monolayer F4-TCNQ is expected to be ca. 3.6 Å [19].

Fig. 1 shows growth and transport properties of as-grown ultra-thin Na$_3$Bi. Fig. 1(a) and 1(b) show the reflection high energy electron diffraction (RHEED) pattern of the α-Al$_2$O$_3$[0001] substrate and ~2 nm Na$_3$Bi film respectively. The clearly defined streaks in Fig. 1(b) are consistent with RHEED taken on ultra-thin Na$_3$Bi [15] and 15 nm Na$_3$Bi grown on Si(111) [20] indicating high-quality epitaxial growth. STM topography in Fig. 1(c) shows ~10 nm terraces with an average thickness of 2 nm (see Supplementary Materials for details). The carrier density and mobility for the nine different as-grown 2 nm Na$_3$Bi samples are shown in Fig. 1(d). In this regime non-local measurements confirm that conduction is dominated by the doped bulk (see below). All as-grown films are *n*-type with carrier density 2.8×10$^{10}$-3.8×10$^{11}$ cm$^{-2}$, and mobility 1000-34000 cm$^2$/Vs. Variations in density and mobility are most likely due to subtle variations in substrate quality and growth conditions, whilst, the *n*-type doping is due to interfacial doping, consistent with 20 nm Na$_3$Bi thin films [17].

We now turn to the bulk magneto-transport properties of as-grown ultra-thin Na$_3$Bi. Fig. 2(a) shows the change in conductivity, $\Delta\sigma_{xx}$, as a function of perpendicular magnetic field for four as-grown samples with different carrier densities. At higher density there is only negative magneto-conductance and as the density decreases there is crossover from negative to positive magneto-conductance at around 0.25 T, consistent with a crossover from WAL to WL.

We fit the data to the full Hikami-Larkin-Nagaoka (HLN) formula (solid lines in Fig. 2(a); see Supplementary Materials for formula and details) to obtain the phase ($l_\phi$) and spin ($l_{SO}$) coherence lengths, and plot the mean free path ($l_e$) (calculated in Supplementary Materials), $l_\phi$, and $l_{so}$ as a function of carrier density in Fig. 2(b). For all films, $l_\phi \gg l_e$ consistent with the quantum diffusive regime. The phase coherence length, $l_\phi$ increases monotonically with doping similar to WSe$_2$ [21]. On the other hand, $l_{so}$ increases as the density decreases, reflecting the decrease in spin-momentum locking (perfect spin-momentum locking corresponds to $l_{so} = 0$). At low density when $l_\phi \sim l_{so}$, the crossover of WAL to WL is expected. Additional experiments probing a *p*-type to *n*-type transition using *in-situ* K-dosing showed a similar transition from WL to WAL (see Supplementary Materials for data).

The coherence length in a conductor is ultimately limited by electron-electron scattering [22] though may be smaller due to other dephasing processes such as electron-phonon scattering or spin-flip scattering from local moments. For electron-electron scattering we expect that the ratio $\frac{l_\phi}{l_e} = \left(\frac{\tau_\phi}{\tau_e}\right)^{1/2} = \left(\frac{4E_F}{k_BT}\right)^{1/2} \left(\frac{1}{ln(\sigma)}\right)^{1/2} \approx \left(\frac{4E_F}{k_BT}\right)^{1/2}$. The electron band of few-layer Na$_3$Bi is well approximated by a massive Dirac system with Fermi velocity $v_F = 1 \times 10^6$ m/s, gap parameter $\Delta = 92$ meV, and effective mass $m^* = \Delta / v_F^2 = 0.016~m_e$, where $m_e$ is the electron mass [15]. Then for our typical carrier densities $n = $ 1-4 $\times 10^{11}$ cm$^{-2}$ and $T = 5$ K we find $\frac{l_\phi}{l_e} \approx$ 12-24. This is in good agreement with the ratio $\frac{l_\phi}{l_e} \approx$ 8-14 observed in Fig. 2(b) (and Fig. S2(b)), implying that the small effective mass and large Fermi energy in $Na_3Bi$ enables long coherence times and phase-coherent transport at relatively high temperature. Further studies of the temperature dependence of the magnetoresistance could confirm whether electron-electron interactions limit the phase coherence.

Bulk $Na_3Bi$ exhibits near perfect WAL due to the perfect spin-momentum coupling expected for a topological Dirac semimetal [17]. The appearance of WL at low carrier density indicates that this topological protection is lost in ultrathin $Na_3Bi$, due to opening of bulk gap, while near perfect WAL is retained at high density far from the gapped region where spin-momentum locking is approximately preserved. The opening of bulk gap is a necessary but not sufficient condition for a 2DTI; it does not indicate whether the resulting insulator is topological or conventional.

We now investigate the transport properties when the Fermi level is within the bulk bandgap. We deplete the as-grown $n$-type doping by depositing high electron affinity (5.24 eV) F4-TCNQ to $p$-type dope thin-film Na$_3$Bi.

Fig. 3 plots (Fig. 3(a)) the local resistance, $R_L$, (Fig. 3(b)) the non-local resistance, $R_{NL}$, and (Fig. 3(c)) their ratio $R_{NL}/R_L$ as a function of F4-TCNQ coverage at temperatures of 5.3-20 K. For a uniform conductor we expect [23] $R_L = \rho_{xx} L/w$ and $R_{NL} = \rho_{xx} e^{-\pi L/w}$. For our geometry, $L/w$ ratio is 2 and hence $R_{NL}/R_L \sim 10^{-3}$. This agrees well with our measurements on the as-grown film where $R_L \sim 10$ k$\Omega$ and $R_{NL}$ is only few ohms. Upon initial deposition of F4-TCNQ, the local resistance increases by two orders of magnitude consistent with the Fermi level moving into the bulk gap. $R_{NL}$ is no longer suppressed and becomes comparable with $R_L$ and $R_{NL}/R_L$ is close to 1 at an F4-TCNQ coverage of 0.020 monolayer (ML). This non-local resistance in the absence of magnetic field has been previously taken as evidence of a 2DTI edge state [10]. We have eliminated the possibility of edge conduction by metallic Bi or Na (see Supplementary Materials) and conclude that conduction occurs through Na$_3$Bi edge modes. For the highest coverages and lowest temperatures, R$_{NL}$ saturates at ~100 M$\Omega$, and $R_{NL}/R_L$ saturates ~ 1, suggesting that when the Fermi level is deep in the bulk gap, the edge current is less affected by bulk leakage, and $R_{NL}$ reflects the edge resistance. The edge resistivity is $\rho_{edge} = R/L = h/e^2 l_{e,edge}$ where $R$ is the resistance of an edge section of length $L$, $h$ is Planck's constant, and $e$ the elementary charge. Accounting for the two current paths, $R_{NL} \sim 100$ M$\Omega$ corresponds to $\rho_{edge} \sim 300$ k$\Omega$/µm and $l_{e,edge} \sim 100$ nm.

Fig. 3(b) shows that $R_{NL}$ is temperature-dependent, becoming larger at lower temperature. However, the temperature dependence appears to mirror the temperature dependence of the ratio $R_{NL}/R_L$; the largest $R_{NL}$ is measured at 5.3 K when $R_{NL}/R_L$ saturates near unity, and lower $R_{NL}$ values coincide with deviation from $R_{NL}/R_L$ ~1 and likely reflects bulk leakage rather than temperature dependence of $\rho_{edge}$. More work is needed to clarify the detailed temperature dependence of $\rho_{edge}$. The temperature dependence of $R_{NL}/R_L$ also prevents us from probing edge transport to temperatures above 20 K, apparently limited by bulk leakage. This temperature is much smaller than the bandgap in ultrathin Na$_3$Bi, suggesting that bulk leakage could be significantly reduced in cleaner samples, realizing higher temperature edge transport.

Fig. 3(d) shows a non-local $I(V)$ curve with around 0.036 ML F4-TCNQ coverage at 10 K when $R_{NL}/R_L$ ~ 1. The $I(V)$ shows two different non-local resistance regimes: at low current, the slope is large ($R_{NL}/R_L$ ~ 1) as electron transport occurs through the edge states. However, at higher current the slope becomes much smaller, corresponding to a decrease in $R_{NL}/R_L$, consistent with some current being shunted through the bulk. The abrupt change in slope indicates that the bulk conduction turns on at a critical edge current/voltage; the critical voltage ~ ±75 mV is of similar order as the bulk bandgap energy hence it is reasonable to expect significant bulk conduction at this bias.

We now demonstrate that the observed edge transport is helical. Fig. 4 shows the longitudinal magnetoresistance (MR) of 2 nm Na$_3$Bi as a function of F4-TCNQ doping. In contrast to the positive MR due to WAL observed for bulk transport (Fig. 2(a) and black line in Fig. 4), in the edge conduction regime a giant negative MR (GNMR) develops, as large as -80% at 0.9 T and nearly independent of the doping (F4-TCNQ coverage). Negative MR in metals is unusual. GNMR has been observed previously in *bulk* Dirac and Weyl semimetals, but only in parallel electric and magnetic fields due to the chiral anomaly [24-27], however the GNMR observed here in ultra-thin Na$_3$Bi is observed in a perpendicular magnetic field. The finite edge resistance in 2DTIs with weak temperature dependence has been explained as due to exchange-mediated scattering from local moments, either atomic-scale defects [1] or electrons localised in puddles [8] due to spatial fluctuations of the Fermi energy, at temperature $T > T_K$, the Kondo temperature. Exchange-mediated scattering from magnetic impurities is a known source of negative MR in metals: Magnetic fields polarize local moments, making spin-flip scattering inelastic and unfavourable once the Zeeman energy exceeds the thermal energy.

We model the edge state magnetoresistance as follows. The edge states of Na$_3$Bi in the presence of a perpendicular magnetic field // **z** are described by the Hamiltonian $H = (\hbar v_F k_x + g_e \mu_B B) \sigma_z$, where **x** is the transport direction parallel to the edge, $g_e$ is the edge spin *g*-factor and the Pauli matrix $\sigma_z$ represents the edge spin operator. The spins are polarised out of the plane and the magnetic field shifts the origin of $k_x$ without opening a gap. We consider scattering off magnetic impurities with spin $\vec{S}$ coupled to the edge spins via a local exchange interactions $J\delta(\vec{r}) \vec{\sigma} \cdot \vec{S}$, whose matrix elements between momentum states are independent of wave vector. The magneto-resistance is calculated using Boltzmann theory in the spirit of Refs.[28,29] but adapted to spin-polarised edge states. Scalar and spin-conserving transitions $\propto \sigma_z S_z$ drop out when the scattering-out and scattering-in terms are both included, eliminating the effect of forward scattering. Single-particle spin-flip terms are determined in the Born approximation $\propto J^2$, keeping terms linear in the impurity density. This yields the following formula for the magneto-resistance:

$$\frac{R(B)-R(0)}{R(0)} = \frac{\langle S_z \rangle}{S}\left[\coth\frac{x}{2} - \frac{\frac{x}{2}}{\left(\sinh\frac{x}{2}\right)^2}\right] \qquad (1)$$

where $x = \frac{g\mu_B B}{k_B T}$, $g$ is the impurity gyromagnetic ratio, $g$-factor, and $\frac{\langle S_z \rangle}{S}$ is given by the Brillouin function, which reflects the polarisation of the impurity spins by the Zeeman interaction. Spin-flip becomes inelastic and the energy cost associated with flipping the impurity spins leads to a negative magnetoresistance. We note that Kondo scattering terms logarithmic in $T$ show up at orders $J^3$ and higher. The Kondo temperature $T_K$ is expected to be vary small for known TI parameters [30], and the good fit to Eq. (1) suggests $T \gg T_K$ here.

Fig. 4 shows a fit of the experimental MR to Eq. (1); the only free parameter is $g = 32.3$. This is in reasonable agreement with $g \sim 20$ inferred from transport measurements on bulk $Na_3Bi$ [25], suggesting the impurities are quantum dot-like "puddles". In Fig.4, we also show the expected MR for a hypothetical non-helical metal, where the only scattering is via exchange interaction, but spin-preserving scattering is allowed; Eq. (1) is modified by an additional prefactor 2/3, i.e. the maximum MR is 2/3 (67%). Generic metals with other scattering processes would show even lower MR. Our best estimate for the fraction of spin-flip scattering in Fig. 4 is 98.1% (see Supplementary Materials). We have observed at higher F4-TCNQ doping that the fraction of spin-flip scattering is reduced (see Supplementary Materials). This is possibly due to the formation of larger puddles with increasing disorder, which scatter via non-spin-flip inelastic processes, or the F4-TCNQ ions themselves acting as a local moments with small $g \sim 2$.

Our results demonstrate how electronic transport experiments can be used to observe the opening of a bandgap due to confinement in a topological Dirac semimetal, and to determine unambiguously whether the resulting insulator is topological or trivial. Our as-grown ultrathin $Na_3Bi$ films are lightly *n*-doped, and the conducting bulk shows a transition from weak anti-localisation to weak localization as the doping is decreased, as expected for a gapped Dirac system. The weak localization behavior contrasts with gapless bulk $Na_3Bi$, which shows only perfect weak anti-localisation, strongly indicating a mixing of the Dirac cones and opening of a gap in ultrathin $Na_3Bi$. When the sample is doped to bring the Fermi level into the bulk bandgap, the non-local resistance signal becomes comparable to the local signal, indicative of edge transport. A giant negative magnetoresistance arises in the edge transport regime due to the suppression of spin-flip scattering by a magnetic field. The field dependence of the resistance shows excellent agreement with theory for a helical edge mode, conclusively demonstrating the topological nature of the edge conductance. We expect that the observed GNMR should be generic to helical edge transport in any 2DTI. Furthermore the GNMR follows a simple universal form with only one materials parameter, $g$. While the GNMR is particularly large in $Na_3Bi$ due to the large $g$, even for $g = 2$ we expect that the GNMR should reach 90% at $B = 3.1$ T and $T = 1$ K. Therefore we expect the combination of non-local transport measurements and magnetoresistance to provide a simple method for unambiguous identification of future 2DTI materials which is not limited to samples exhibiting ballistic transport over long distances.

Acknowledgments

We acknowledge Jian-Hao Chen for valuable discussions. The authors acknowledge support of the ARC Centre of Excellence FLEET (CE170100039). M. T. E. is supported by ARC DECRA fellowship (DE160101157). This work was performed in part at the Melbourne Centre for Nanofabrication (MCN) in the Victorian Node of the Australian National Fabrication Facility (ANFF).

Author contributions

All the authors participated in designing the experiments. C.L. grew the samples and carried out the electronic transport measurements with assistance from M.T.E. C.L. analysed the data, and prepared the draft manuscript. D.C provided theoretical interpretation of the unique giant magnetoresistance data. M.T.E. and M.S.F. provided further analysis of the data, and edited the manuscript.

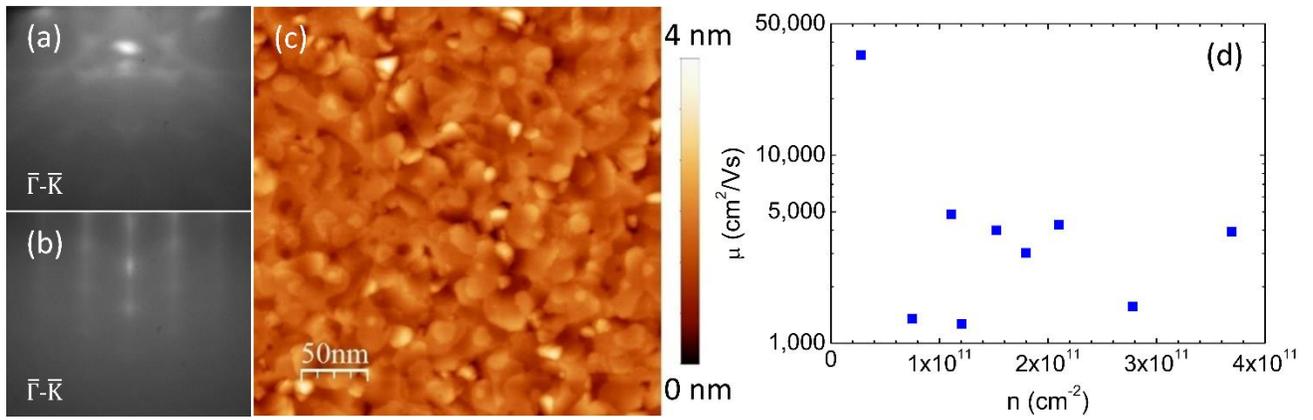

FIG. 1. Growth of 2 nm Na$_3$Bi on sapphire. RHEED patterns for (a) the α-Al$_2$O$_3$[0001] substrate prior to growth and (b) after growth of 2 nm Na$_3$Bi. (c) Large-area (300 nm x 300 nm) topographic STM image (bias voltage $V = 4$ V and tunnel current $I = 100$ pA) of 2 nm Na$_3$Bi on α-Al$_2$O$_3$[0001]. (d) Carrier mobility as a function of Hall carrier density at 5.3 K.

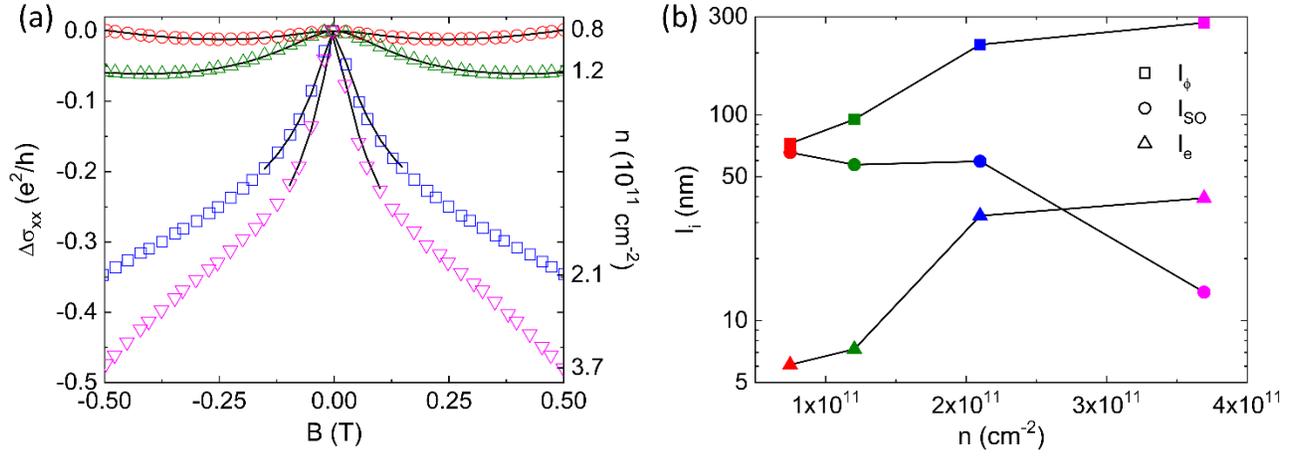

FIG. 2. Magneto-transport properties of four as-grown 2 nm Na$_3$Bi films at a temperature of 5.3 K. (a) Change in magneto-conductivity, $\Delta\sigma_{xx}$ as a function of perpendicular magnetic field $B$ for four films with carrier densities indicated on right axis. Solid lines are fits to the H.L.N formula (see SI for details). (b) Extracted phase coherence length $l_\phi$ (squares), spin-scattering length $l_{so}$ (circle) and elastic-scattering length $l_e$ (triangle) of each film in (a) as a function of carrier density. The colours in (b) match those used in (a).

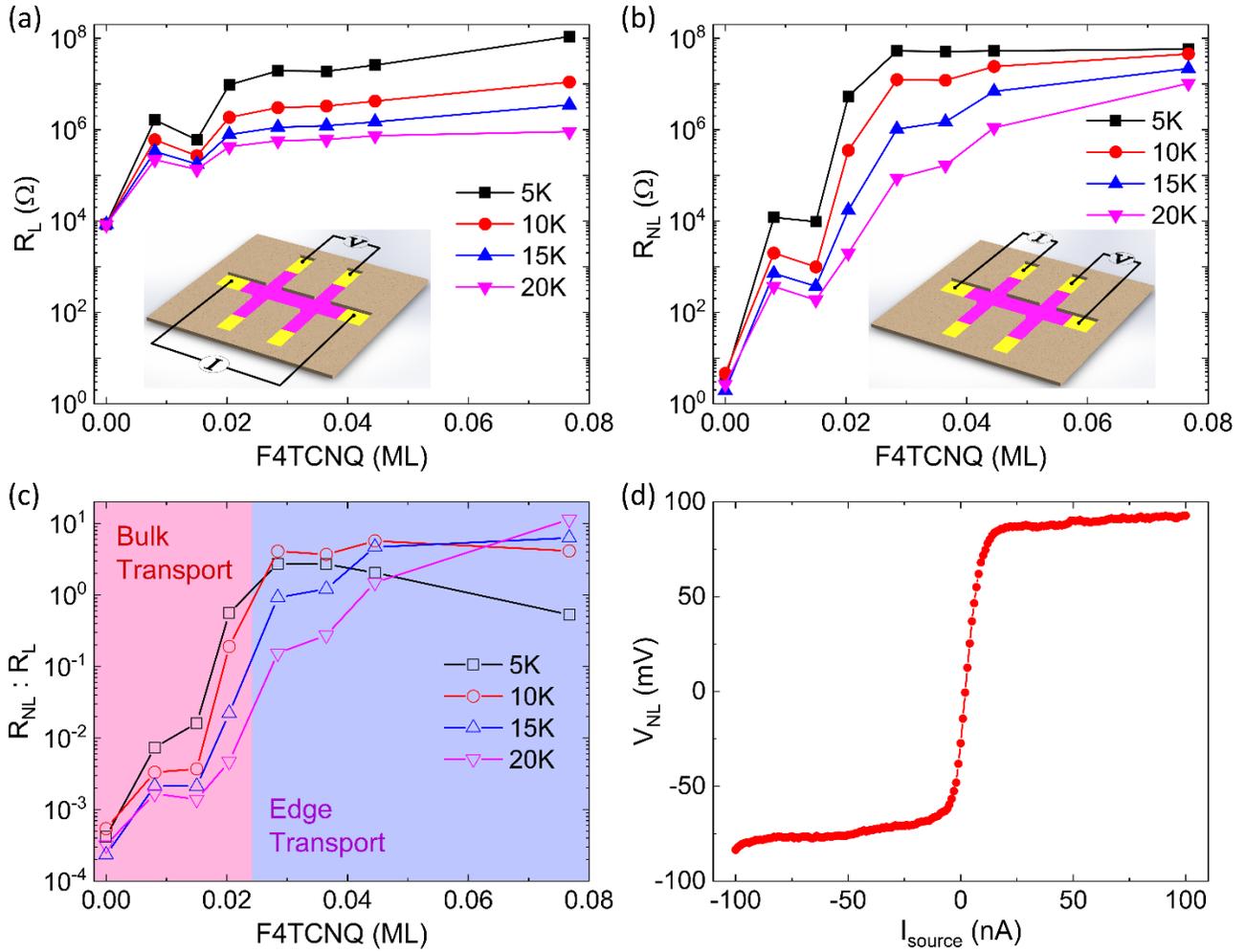

FIG. 3. Local and non-local resistance measurement as a function of F4-TCNQ coverage. The local resistance $R_L$ (a) and non-local resistance $R_{NL}$ (b) and the ratio $R_L : R_{NL}$ (c) are shown as a function of F4-TCNQ thickness at temperatures indicated in legend. The film is made less *n*-type with increasing F4-TCNQ thickness. (d) Non-local *IV* curve of non-local resistance measurements at 10 K with 0.036 ML F4-TCNQ coverage. Insets in (a) and (b) show the geometries used for local and non-local measurements respectively.

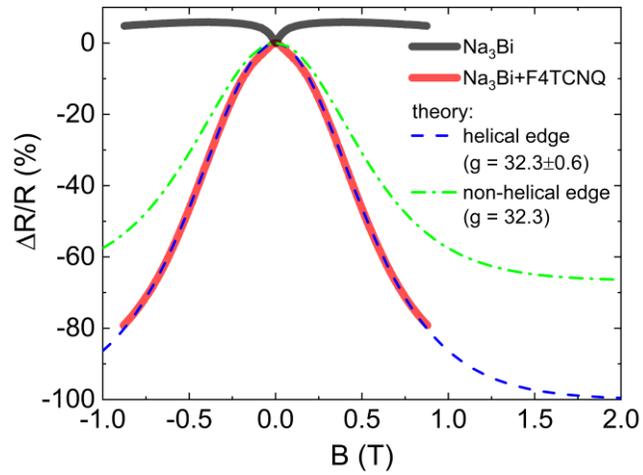

FIG. 4. Resistance as a function of magnetic field at $T = 5.3$ K for bulk-conducting film (as grown, black line) and edge-conducting film (0.015 ML F4-TCNQ coverage, red line). Fits to theory for exchange-mediated scattering for helical and non-helical edges are shown as blue dashed and green dot-dashed lines, respectively.

Supplemental Material for

Signatures of Helical Edge Transport in Millimetre-Scale Thin Films of Na$_3$Bi

Chang Liu[1,2,3], Dimitrie Culcer[3,4], Mark T. Edmonds[1,2,3*], Michael S. Fuhrer[1,2,3*]

[1]School of Physics and Astronomy, Monash University, Victoria 3800 Australia

[2]Monash Centre for Atomically Thin Materials, Monash University, Victoria 3800 Australia

[3]ARC Centre of Excellence in Future Low-Energy Electronics Technologies, Monash University, Victoria 3800 Australia

[4]School of Physics, University of New South Wales, Sydney, New South Wales 2052 Australia

\* mark.edmonds@monash.edu and michael.fuhrer@monash.edu

**Scanning tunneling microscopy topography**

Fig. S1 presents height and roughness analysis of Fig. 1(c) of the main manuscript. Fig. S1(b) shows a height histogram of the topography image, with a predominant peak at ~1.5 nm, measured relative to rare pinholes which we identify as the $Al_2O_3$ surface, with two slight plateaus at ~1 nm and ~2 nm. A monolayer (half unit cell) of $Na_3Bi$ corresponds to 0.48 nm, hence Fig. S1(b) is consistent with a predominantly tri-layer $Na_3Bi$ film, with co-existing regions of bi-layer and quad-layer $Na_3Bi$. Fig. S1(d) shows the film height along the green line of Fig. S1(a) which passes through a pinhole to the substrate, with most plateaus around 1.5 nm.

Fig. S2(a) is an STM topography (500 nm x 250 nm) taken from a different sample grown under nominally the same growth conditions. This is comparable with the high quality ultra-thin $Na_3Bi$ film grown on Si [1]. Here, we can see the grain size is around 25 nm in two different areas, with the bright regions corresponding to bi-layer, and the dark regions mono-layer. The height histogram shown in Fig. S2(b) has two clear peaks ~0.5 nm and ~1.0 nm, consistent with mono-layer and bi-layer regions given a monolayer of $Na_3Bi$ corresponds to 0.48 nm. Fig. S2(c) and S2D are two height line profiles representing the green and red line traces shown on Fig. 2(a) respectively.

The variation in film quality between two nominally identically grown films is most likely due to small variations in the sapphire substrates influencing the growth quality. To achieve the highest quality ultra-thin $Na_3Bi$ that will enable edge transport at higher temperature will therefore rely on improving the preparation of sapphire substrates in order to reliably reproduce films of the quality shown in Fig. S2.

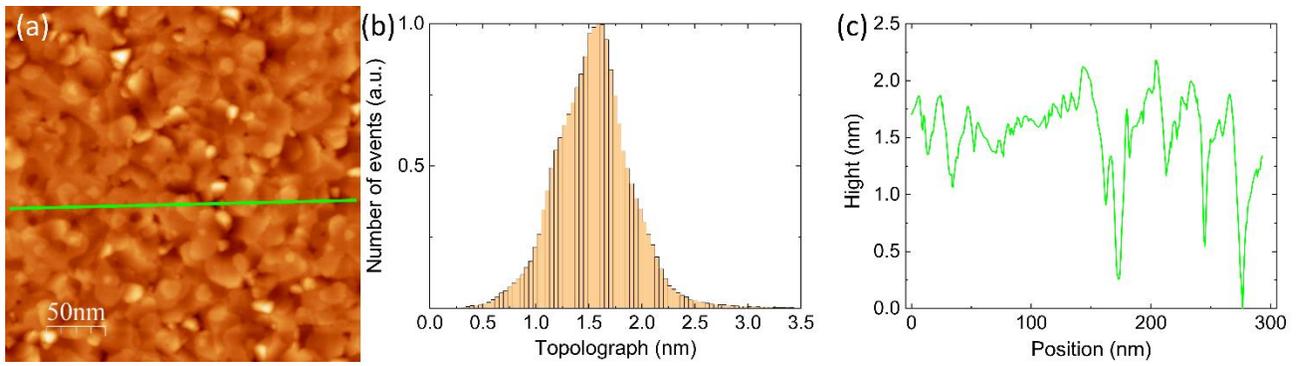

FIG. S1. Thickness analysis of STM image Fig. 1(c). (a) Topographic STM image (300 nm x 300 nm with bias voltage $V = 4$ V and tunnel current $I = 100$ pA) of Fig. 1(c) with respective topography histogram (b). The crystal height along the green line in (a) shows in (c).

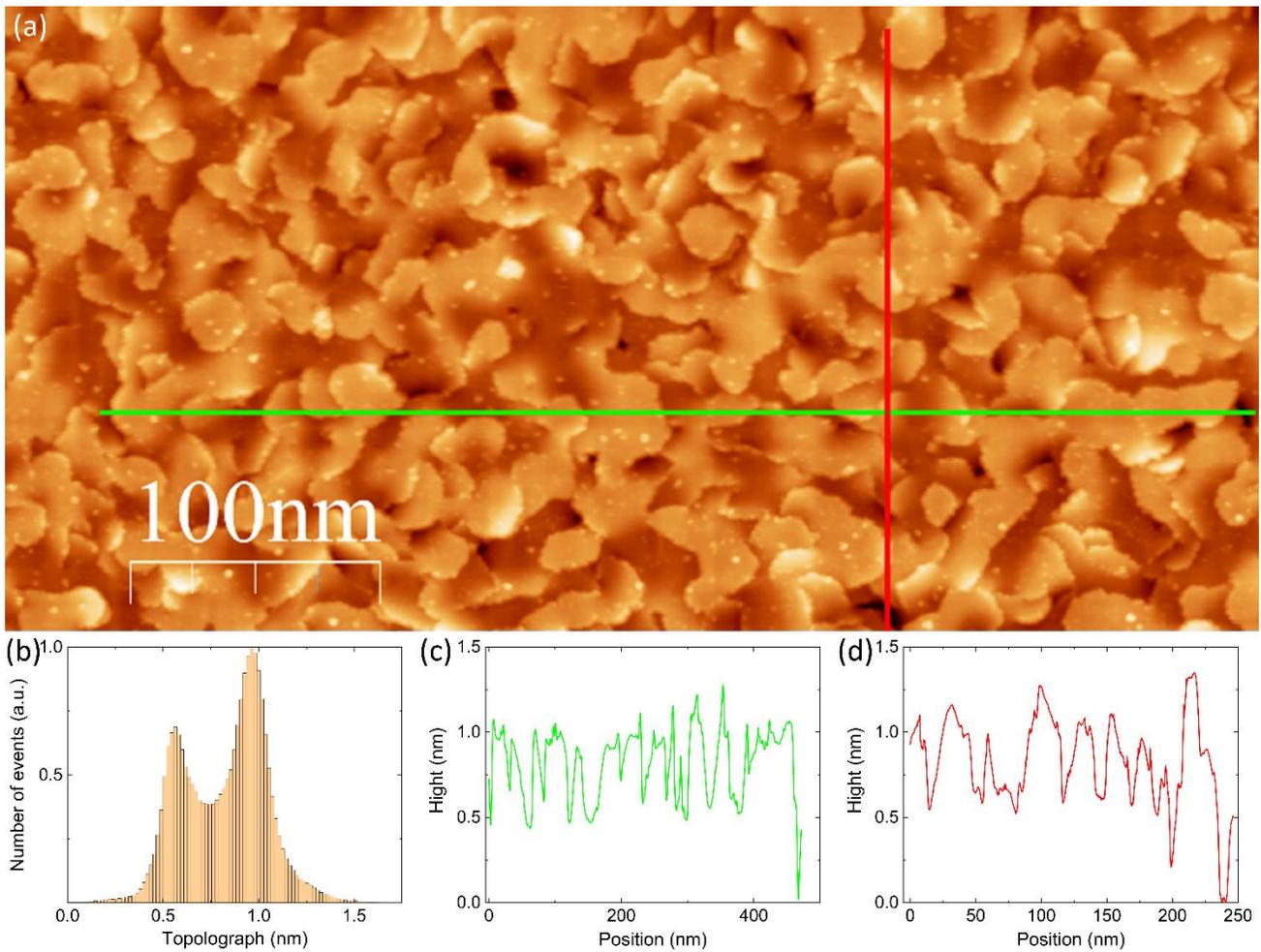

FIG. S2. STM image of an ultrathin area. (a) Topographic STM image (bias voltage $V = 3$ V and tunnel current $I = 100$ pA) of a large flat area (500 nm x 250 nm) on another film with the respective histogram in (b). The crystal height along the green and red lines in (a) show in (c) and (d).

**Magnetoresistance further analysis and discussion**

1. Magnetoresistance measurement

The magnetic field-dependent longitudinal and Hall resistances $R_{xx,m}(B)$ and $R_{xy,m}(B)$ were measured using the conventional Hall-bar technique. To eliminate small contributions of the other resistance components, we symmetrize data as $R_{xx}(B) = [R_{xx\_m}(B) + R_{xx\_m}(-B)]/2$ and $R_{xy}(B) = [R_{xy\_m}(B) - R_{xy\_m}(-B)]/2$. We define the longitudinal magnetoresistance, MR(%) = $100 \times [R_{xx}(B) - R_{xx}(0)]/R_{xx}(0)$. We calculated the 2D Hall carrier density $n = \frac{1}{e}\frac{dB}{dR_{xy}}$ and Hall mobility $\mu = \frac{1}{ne R_{xy}}$. Here, the *n*-type carriers are defined as positive, and the *p*-type carriers as negative.

2. Extracting mean free path

The momentum relaxation length (or the mean free path) can be extracted from our classical transport data, using $l_e = \frac{\hbar}{e}\mu\sqrt{2\pi n}$ where *e* is the elementary charge and $\hbar$ is the reduced Planck constant. This allowed us to fit the magnetoconductivity data to the Hikami-Larkin-Nagaoka formula (discussed below) with only two free parameters, instead of three.

3. Hikami-Larkin Nagaoka formula

To understand the quantum interference correction to the conductivity, we need to study the phase coherence ($l_\phi$), spin-orbit ($l_{so}$) and elastic ($l_e$) scattering lengths and their respective characteristic fields, $B_i$, where $B_i = h/(8\pi e l_i^2)$ (*i*=$\phi$, *so*, *e*). We define the magneto-conductivity as $\Delta\sigma_{xx}(B) = \frac{\sigma_{xx}(B) - \sigma_{xx}(0)}{e^2/h}$ in unit of $e^2/h$ where *e* is elementary charge, *h* is Planck constant and $\sigma_{xx}(B) = \frac{R_{xx}(B)}{R_{xx}(B)^2 + R_{xy}(B)^2}$ or $\sigma_{xx}(B) = \frac{1}{R_{xx}(B)}$ when $R_{xx}(B) \gg R_{xy}(B)$. The magneto-conductance data is then fitted to the Hikami.Larkin.Nagaoka (H.L.N) formula [2,3]:

$$\Delta\sigma_{xx}(B) = -\frac{1}{\pi}\left\{\frac{1}{2}\left[\Psi\left(\frac{1}{2}+\frac{B_\phi}{B}\right) - \ln\left(\frac{B_\phi}{B}\right)\right] + \left[\Psi\left(\frac{1}{2}+\frac{B_{so}+B_e}{B}\right) - \ln\left(\frac{B_{so}+B_e}{B}\right)\right] \right.$$
$$\left. - \frac{3}{2}\left[\Psi\left(\frac{1}{2}+\frac{\frac{4}{3}B_{so}+B_\phi}{B}\right) - \ln\left(\frac{\frac{4}{3}B_{so}+B_\phi}{B}\right)\right]\right\} \quad (S1)$$

where $\Psi$ is the digamma function. The $B_e$ term is fixed using the value determined from our classical transport measurements.

4. Magneto-transport evolution of a p-type film with K-dosing

Interestingly, one of the films was found to become $p$-type after the superconducting magnet quenched (causing a rapid heating of the cryostat and sample stage to 50 K), resulting in the desorption of vacuum species such as CO, causing the $p$-type doping. The magneto-conductance of the now $p$-type film is plotted as a red trace in Fig. S3(a), and shows entirely weak localization. This then allowed us to study the evolution of the magneto-transport as a function of doping via *in-situ* K-dosing as shown in Fig. S3(a). A clear transition from pure WL (red-circle) at low $p$-type doping to a crossover of WL and WAL (green-triangle) at low $n$-type doping and to pure WAL at higher $n$-type doping level is observed with increasing K-dosing. The HLN analysis $l_\phi$ and $l_{SO}$ is shown in Fig. S3(b), and display a similar carrier density dependence to that observed for the as-grown films *i.e.* with increasing carrier density, $l_\phi$ is increases whilst $l_{so}$ decreases. Again, the crossover from WL to WAL occurs when $l_\phi \sim l_{so}$, as expected.

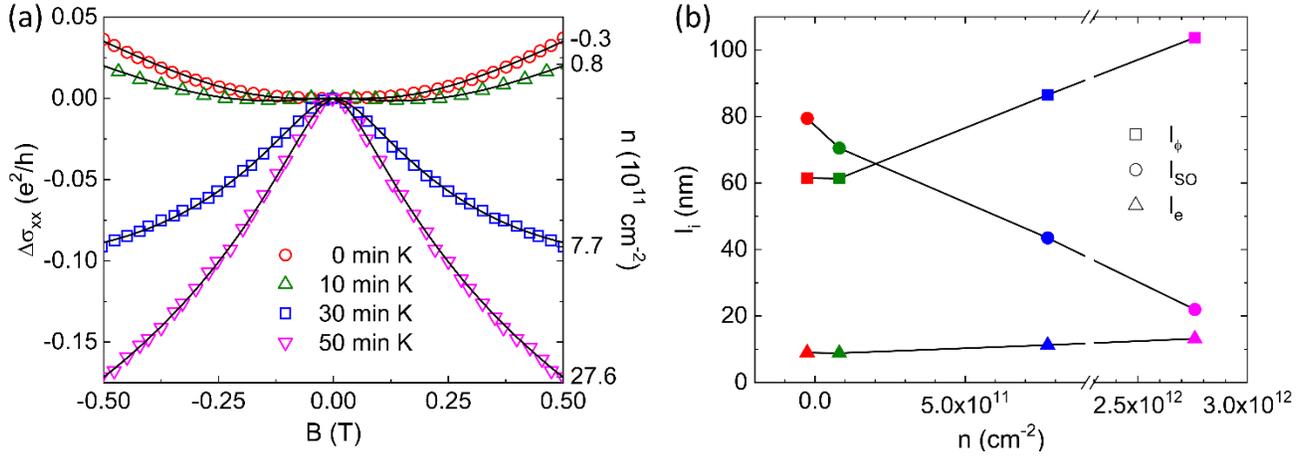

FIG. S3. Magneto-transport properties of 2 nm Na$_3$Bi with K-doping. (a) Evolution of magneto-conductance, $\Delta\sigma$, of an initially $p$-type (shown as negative carrier density) film with K-doping time. Solid lines are fits to the HLN formula. (b) Extracted phase coherence length $l_\phi$ (squares), spin-scattering length $l_{so}$ (circle) and elastic-scattering length $l_e$ (triangle) as a function of Hall carrier density which increased with K-doping time (color). The colours in (b) match those used in (a).

**Possibility of conduction through metallic 1D Na or Bi**

One possible alternative to edge transport in ultra-thin Na$_3$Bi, is that at the edge of the sample exists a pure Na or Bi nanowire (as a result of the Na$_3$Bi film being grown through a shadow mask) which would display 1D metallic conduction. To rule out this possibility we grew two separate films; one consisting of only Bi and one consisting of only to determine whether either Bi or Na grow separately on the sapphire substrate. These growths were performed at the same substrate temperatures and conditions used for the Na$_3$Bi growth. For both the Bi film and Na film the resistance (measured locally or non-locally) is above G$\Omega$ (which is significantly larger than 100 M$\Omega$ measured for edge transport in the Na$_3$Bi ultra-thin films). This allows us to rule out that the non-local edge state transport is caused by pure Na or Bi metal nanowires.

**Spin-flip and spin-preserving scattering**

The magneto-resistance due to pure spin-flip scattering is predicated by Eq. (1) (in main manuscript). For a hypothetical non-helical metal, where the only scattering is via exchange interaction, but spin-preserving scattering is allowed the expected MR is given by:

$$\Delta R/R = -\frac{1}{S+1}\frac{\langle S_z \rangle}{S}\left[\coth\left(\frac{1}{2}x\right) - \frac{\frac{1}{2}x}{\sinh^2\left(\frac{1}{2}x\right)}\right] \quad (S2)$$

where $S = \frac{1}{2}$ is the impurity spin and $x = g\frac{\mu_B}{k_B T}B$. It is clear Eq. (S2) is the same as the Eq. 1 but with an extra term, $\frac{1}{S+1} = \frac{2}{3}$. Thus, the maximum MR limit is 2/3 for this hypothetical non-helical metal.

We can generalize the expression to an arbitrary metal with a fraction of scattering due to spin-flip processes $A$:

$$\Delta R/R = -A\frac{\langle S_z \rangle}{S}\left[\coth\left(\frac{1}{2}x\right) - \frac{\frac{1}{2}x}{\sinh^2\left(\frac{1}{2}x\right)}\right] \quad (S3)$$

We have studied the MR for ultrathin Na$_3$Bi films doped into the bulk insulating regime at different F4-TCNQ coverages, as shown in Fig. S4. We find that the MR is reduced at higher F4-TCNQ coverages. If we assume that the gyromagnetic ratio remains unchanged ($g = 32.3$), we can fit the curves to Eq. S3 with a single parameter $A$, as shown in Fig. S4(a). The fits are reasonably good, and show that $A$ varies from near 100% at low doping, to 64% at high doping. If we allow both $A$ and $g$ to vary as fit parameters, we find the best fits to the data as shown in Fig. S4(b). At low-doping (as shown in Fig. 4 of main text) the best fit is $A = 0.981 \pm 0.002$ and $g = 32.9 \pm 0.7$. Thus our best estimate of the fraction of spin-flip scattering is 98.1%. Fits to the data at higher F4-TCNQ coverages show little variation in g, but a significant variation in *A*, consistent with our hypothesis that *g* is independent of doping.

The remaining fraction of the zero-field resistance 1-*A* may represent inelastic scattering induced by the disorder potential [4] or by puddles without moments [5], both of which should increase at higher F4-TCNQ coverages due to higher disorder, and show weak magnetic field dependence. Alternatively, the fraction of scattering 1-*A* may be due to spin-flip scattering by an additional population of local moments with low gyromagnetic ratio, for example the adsorbed F4-TCNQ- ions which should have $g \sim 2$. The fit to the spin-flip scattering model for two populations of moments with fraction *A* having $g \sim 32$ and fraction 1-*A* having $g \sim 2$ is indistinguishable from the fit to Eq. S3 in the field range studied, so we cannot distinguish between these two scenarios. More experiments at variable temperature and a larger magnetic field range would help to clarify the scattering mechanisms in the edge state, however, are not possible in our UHV transport setup which has a fixed temperature ($T = 5.3K$) and field range up to 1 T.

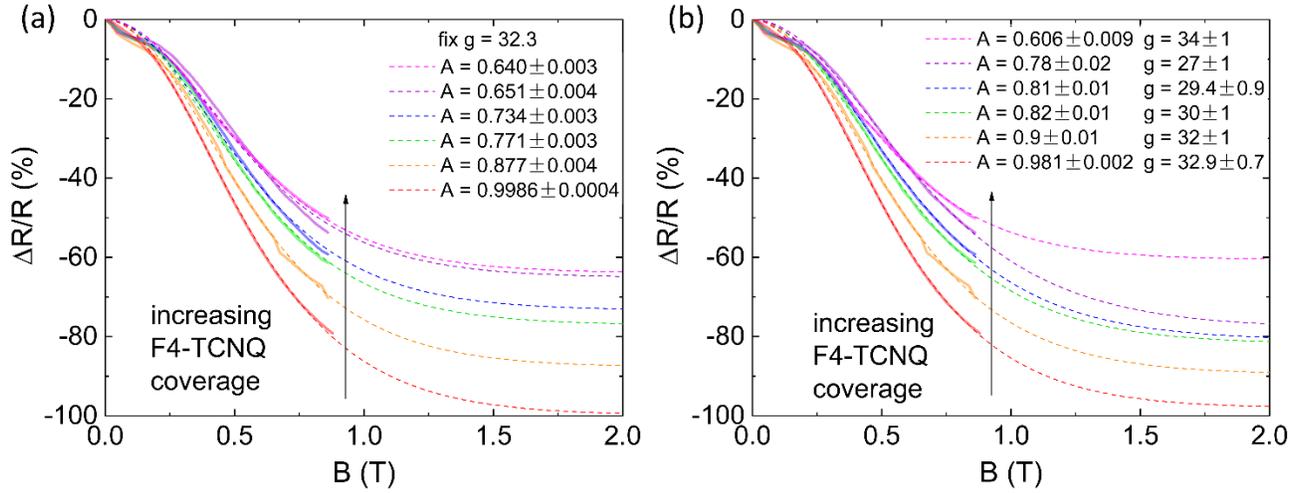

FIG. S4. Resistance as a function of magnetic field at $T = 5.3$ K for edge-conducting film with increasing F4-TCNQ coverage: 0.015 ML (red same as the red curve in Fig. 4), 0.020 ML (orange), 0.028 ML (green), 0.036 ML (blue), 0.044 ML (violet) and 0.077 ML (magenta). The dashed line is the fitting to Eq. S3. For (a) there is only one free parameter, $A$, with fixed $g = 32.3$, whilst (b) has both $A$ and $g$ as fitting parameters.